\documentclass[12pt]{article}
\usepackage[english]{babel}
\usepackage{epsfig,macr,slett}
\begin{document}
%\maketitle

\begin{titlepage}
\begin{flushright}
  DESY 01-096 \\
\end{flushright}

\vskip 1 cm
\begin{center}
  {\Large\bf Testing perturbation theory on the $\nf=0$ static quark potential}
\end{center}
\vskip 1.0cm
\begin{center}
{\large Silvia Necco and Rainer Sommer}
\vskip 0.8cm
DESY, Platanenallee 6, D-15738 Zeuthen, Germany
\vskip 0.8cm
 Silvia.Necco@desy.de
\vskip 0.1cm
 Rainer.Sommer@desy.de
\end{center}
\vskip 2.5ex
{\bf Abstract}
\vskip 0.7ex

The perturbative expansion of static force and potential is reanalyzed
concerning its practical applicability. A well behaved 
perturbative prediction is given
by the integration of the renormalization group equation
for the coupling $\alphaqq(\mu=1/r)=(\Cf)^{-1}r^2F(r)$. 
Since the
Lambda-parameter of the $\nf=0$ theory is known from \cite{mbar:pap1},
the perturbative prediction contains no free parameter. 
It is confronted with recent non-perturbative results.
For $\alpha \lesssim 0.3$ where the truncation error of the 
perturbative expression
is naively estimated to be moderate,
it is really quite accurate
and large ``non-perturbative 
terms'' are excluded. 

  \vfill

\begin{flushleft}
  DESY 01-096\\
  September 2001
\end{flushleft} 
\eject

\end{titlepage}

%%% Local Variables: 
%%% mode: latex
%%% TeX-master: "pap1"
%%% End: 

\def\lambdas{{\Lambda_{S}}}
\def\Lambdas{{\Lambda_{S}}}
\def\alphas{{\alpha_{S}}}
\def\alphasp{{\alpha_{S'}}}
\def\lambdasp{{\Lambda_{S'}}}
\def\Lambdasp{{\Lambda_{S'}}}
\newcommand{\ssect}[1]{\noindent {\bf  #1.}}

%\section{Introduction}

\ssect{1} The perturbative expansion in the coupling $\alpha_S$ 
is the most important theoretical 
tool for analyzing strong interaction effects in 
high energy scattering experiments. Since the coupling 
decreases with increasing momentum transfer, 
perturbation theory becomes accurate in the high
energy regime. In practical applications, in particular
in the determination of the running coupling itself,
it is tempting to apply the perturbative series already
where the coupling is not so small. Popular
examples are the determinations of $\alpha_S$ from
$\tau$-lepton decays and the perturbative 
evolution of deep inelastic scattering structure 
functions starting at a renormalization point below 
1~GeV. 
In order to establish the applicability of perturbation theory,
it would be very desirable to study such processes systematically
as a function of the energy, but either they involve a fixed
energy ($\tau$ decays%\cite{Rtau}
, hadronic Z-decays%\cite{RZ}
) 
or the precision is not sufficient 
over a larger energy range (deep inelastic scattering, $e^+e^-$
total cross section%\cite{Ree}
, Adler function%\cite{Adler:Fred}
).
Our main phenomenological 
test of perturbation theory therefore is the overall consistency
of the determinations of $\alpha_S$ from different processes 
(see \cite{alpha:bethke} and references therein).

Complementary information may be obtained from suitable observables 
computed as a function of Euclidean external momenta (or distance) 
using lattice QCD.
Here an important
limitation is implied by the necessarily finite lattice spacing $a$ and 
the corresponding momentum cutoff $\rmO(1/a)$. For QCD with dynamical 
quarks (and in large volume), one currently 
reaches $a^{-1}\approx 2\GeV$. This limitation may be overcome by 
considering a finite size effect as the physical 
observable~\cite{alpha:sigma} which defines a renormalized 
coupling\footnote{In 
         finite volume, with no other scale involved but the
         size of the space-time itself, one may of course
         keep the lattice spacing small compared to this scale.}.
In the \SF framework, the method has been completely developed
and results for the running coupling are 
available~\cite{alpha:SU2,alpha:SU3,mbar:pap1,alpha:lett}, including
in particular the Lambda parameter expressed in terms of the
low energy scale $\rnod \approx 0.5\,\fm$ \cite{pot:r0} in the theory without 
quarks \cite{mbar:pap1} (pure Yang-Mills theory). These results 
refer to the continuum limit, reached by a controlled
extrapolation from finite $a$. In this theory
the scale dependence
of the \SF coupling, $\alphaSF(\mu)$, is in remarkable 
qualitative agreement
with perturbation theory 
for $\alpha<0.3$ and for 
$\alpha<0.2$ the 3-loop expression describes $\alphaSF(\mu)$
within  better than 2\%.
It is an interesting question, 
whether the coupling in this scheme is a special case or
whether this is a more ``general property of the theory''.

If one restricts oneself to $\nf=0$, also observables in large 
volume but still relatively small distances may be computed employing
very large lattices ($64^4$). We have done so for the potential between
static quarks in the fundamental representation, reaching
distances of $r \approx 0.05\fm$ with small discretization 
errors and in fact extrapolating to the continuum for
$r \gtsim 0.1\fm$\cite{pot:silvia1}. 
In perturbation theory, the potential has
been computed to two loops \cite{Fischler:1977yf,Billoire:1980ih,Peter:1997me,Schroder:1998vy,Melles:2000dq}, but at the same time the 
usefulness of perturbation theory even at distances as short as
$0.1\fm$ has been doubted \cite{Peter:1997me,Schroder:1999sg}. 
We shall explain below that this 
is a question of a suitable renormalization scheme. When
the most natural scheme (defined in terms of the force)
is adopted, 
perturbation theory is well behaved at such distances
and it is interesting to compare perturbation theory to 
the non-perturbative force. This is a rather stringent test of
perturbation theory since besides the scale dependence of the
coupling, its absolute value is predicted by perturbation theory
(we remind the reader that the Lambda parameter is known).

A previous exploratory investigation, concentrated on the question, 
whether there are ``large non-perturbative'' terms in the potential 
at short distances \cite{pot:bali99} as they had been argued to exist
\cite{pot:npterms}. 
It is not easy to give a definition of 
``large non-perturbative term''. We assume here that roughly the 
following is meant by this statement. 
\begin{itemize}
\item[(i)] A certain quantity, here the potential $V(r)$, is considered 
in a region where its perturbative expansion looks well behaved, i.e.
the $n$-loop contribution is a small correction and significantly 
smaller than the $(n-1)$-loop contribution (unless the latter is 
accidentally small itself).
\item[(ii)] The difference between the full non-perturbative observable 
and the truncated perturbative series is much larger than the 
last term in the series.
\end{itemize}
With such a definition, necessarily somewhat phenomenological in
character,
we shall demonstrate below that there are definitely no 
large non-perturbative terms in the potential. To the contrary, 
perturbation theory works remarkably well where the criterion 
(i) is satisfied.

In the following, we first investigate the perturbative expressions
and find that a stable perturbative prediction satisfying $(i)$
in an accessible region of $r$ is given
by the integration of the renormalization group equation
for the coupling $\alphaqq(\mu)$.
We then compare perturbation theory to our
numerical results, considering also the direct 
relation between the \SF coupling $\alphaSF$ and $\alphaqq(r)$.
For completeness we also show the potential itself compared to 
perturbation theory.
\vspace{0.5cm}
 
%%% Local Variables: 
%%% mode: latex
%%% TeX-master: "pap2"
%%% End: 

\ssect{2}
In single scale problems, such as the static potential
depending only on the separation $r$, the 
best perturbative prediction is expected to be
the integration of the renormalization group equation. 
This is equivalent to defining a physical renormalized 
coupling, often denoted effective charge  
\cite{alpha:Grunberg_lett,alpha:Grunberg_pap}. In particular
the coupling $\alphaqqbar(\mu)$
may be defined through
\be \label{e_alphaqqbar}
F(r)={\rmd V \over \rmd r}=\casim{\alphaqqbar(\mu) \over r^2}\,,\;\mu=1/r.
\ee
(It will become clear below, why we here consider the force rather than 
the potential.)
The running of the coupling defines the associated $\beta$-function,
\be\label{e_beta}
\mu {\rmd \over \rmd \mu}\gbar = \beta(\gbar)\,,\;\gbar=(4\pi\alpha)^{1/2}
\ee
with a perturbative expansion 
\bes\label{e_beta_pert}
 \beta(\gbar) &{\raisebox{-.3ex}{$\stackrel{\gbar \rightarrow 0}{\sim}$}}
                    &  -\gbar^3 
                     \{ b_0 + b_1 \gbar^2 + b_2 \gbar^4 + \ldots \} \\
              && b_0=\frac{1}{(4\pi)^2}\left(11 - \frac{2}{3}\nf\right),
              \, b_1=\frac{1}{(4\pi)^4}\left( 102 - \frac{38}{3} \nf\right).
\ees
The solution of \eq{e_beta},
\be\label{e_lambda}
\lambdas=\mu(b_{0}\gbar^{2})^{-b_{1}/(2b_{0}^2)}
\rme^{-1/(2b_{0}\gbar^{2})}\exp
\bigg\{-\int_{0}^{\gbar}dx
\left[\frac{1}{\beta(x)}+\frac{1}{b_{0}x^{3}}-
\frac{b_{1}}{b_{0}^{2}x}\right]\bigg\},
\ee
relates the coupling $\gbar=\gbar(\mu)$ to the Lambda-parameter.
This general expression turns into a 
perturbative one by inserting the expansion  \eq{e_beta_pert} for the
$\beta$-function.
Truncating in \eq{e_beta_pert} after the term 
$b_{n-1}$ and solving \eq{e_lambda} (numerically) for $\gbar$ at given
$\mu$ (in units of $\Lambda$) defines the $n$-loop RG solution for the 
coupling. In contrast to the frequently used expansion of $F(r)$ 
(or $V(r)$) in terms of $\alphaMSbar(\mu)$, one does not need to choose
the scale $\mu$.
For the \SF coupling, this perturbative prediction ($n=3$) 
has been shown to be 
rather accurate for $\alpha<0.3$ by comparison to non-perturbative 
results~\cite{alpha:SU3,mbar:pap1}.

In order to obtain $\alphaqq$ from  \eq{e_lambda} we need to insert 
the  Lambda-parameter in this scheme.
We start from 
\be \label{e_lamrnod}
\Lambda_{\MSbar}^{(0)} \rnod=0.602(48)
\ee
referring to the case $\nf=0$ and extracted at sufficiently high 
scale $\mu$ where the perturbative error is negligible \cite{mbar:pap1}.
With the coefficient 
$c_0$ (known from \cite{Fischler:1977yf,Billoire:1980ih}) in the expansion 
\be
  \alphaqqbar(\mu) = \alphaMSbar(\mu) + c_0 \alphaMSbar(\mu)^2 + 
                      c_1 \alphaMSbar(\mu)^3 + \ldots 
\ee
we then relate $\Lambda_{\qqbar}$ to $\Lambda_{\MSbar}$ via
\be
  \Lambda_{\qqbar} = \Lambda_{\MSbar}\, \rme^{c_0/(8\pi b_0)}\,.
\ee
For convenience
the ratio of Lambda-parameters is listed in \tab{t_lambda_b2} together 
with that ratio for
other schemes.

\begin{figure}[ht]%\label{f_alpha_pert}
\begin{center}
\includegraphics[width=9cm]{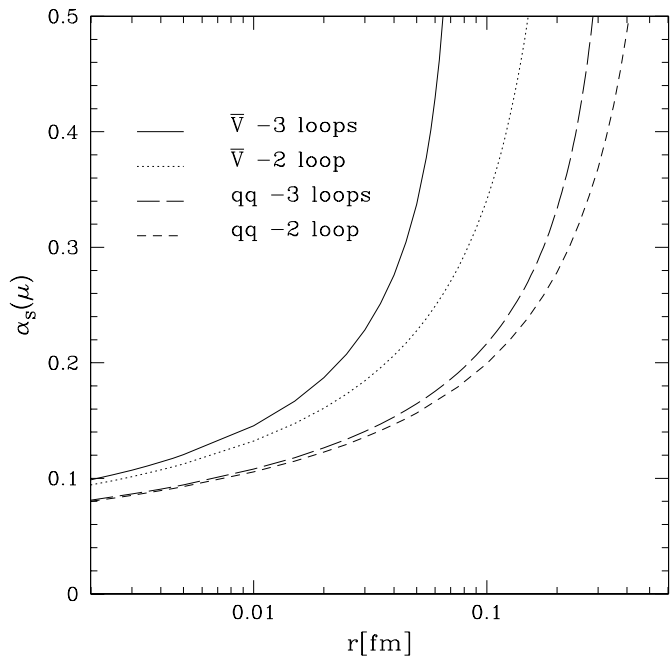}
\end{center}\vspace{-0.8cm}
\caption{\label{f_alpha_pert} \footnotesize 
Running couplings obtained by integration of the 
RG with truncation of the $\beta$-functions at 2- and 3-loop 
and with  
$\Lambda_{\MSbar}^{(0)}=238\MeV$ \protect\cite{mbar:pap1}.} 
\end{figure}

The coupling $\alphaqqbar$ from 2- and 3-loop RG is illustrated in 
\fig{f_alpha_pert}, using the central value 
$\Lambda_{\MSbar} \rnod=0.602$ (the 
8\% overall uncertainty  of this number 
corresponds to a common small horizontal shift of all curves
in the figure).
The perturbative expansion appears 
quite well
behaved up to distances $r\sim 0.25 \fm$. At $r\sim 0.2 \fm$
one would expect the 3-loop curve to have an accuracy of about 10\%.
Since the force is completely equivalent to $\alphaqqbar$
it is given with the same relative accuracy; the potential
may be obtained  
by integration of the force.
%%%%%%%%%%%%%%%%%%%%%%%%%%%%%%%%%%%%%%%%%%%%%%%%%%%%%%%%%%%%%%%%%%%%%%%%%%%%%%%%%%%%%%%%%%%%%
%%%%%%%%%%%%%%%%%%%%%%%%%%%%%%%%%%%%%%%%%%%%%%%%%%%%%%%%%%%%%%%%%%%%%%%%%%%%%%%%%%%%%%%%%%%%%

Alternatively one may define couplings $\alphavbar$ and $\alphav$ through the 
potential 
\be \label{e_alphavbar}
  V(r) = -\casim{\alphavbar(\mu)\over r}\,,\;\,\mu=1/r
\ee
and its Fourier transform
\be
  \tilde{V}(Q) = -4\pi \casim{\alphav(Q) \over Q^2}\,.
\ee
We note, however, that the 3-loop coefficients $b_2$ 
are larger 
in these cases (see \tab{t_lambda_b2}), in particular
in the $\overline{V}$ scheme. As a consequence, the 
difference between the 2-loop and the
3-loop running coupling in this scheme 
is only small at very short distances and this
perturbative expansion appears to be applicable only up to 
$\alpha \sim 0.15$. This is also illustrated in \fig{f_alpha_pert}.
%%%%%%%%%%%%%%%%%%%%%%%%%%%%%%%%%%%%%%%%%%%%%%%%%%%%%%%%%%%%%%%%%%%%%%%%%%%%%%%%%%%%%%%%%%%
\begin{table}[ht]
\begin{center}
\begin{tabular}{ c |  c  c  c  c c }
\hline\\[-1ex]
          &  $S\,=\,$ & $\qqbar$  & $V$  & $\overline{V}$  &  $\rm SF$  \\[1ex]
\hline\\[-1ex]
$\Lambda_{S}/\Lambda_{\MSbar}$ & &  $\exp(\gamma-35/66)$ & $\exp(31/66)$ & 
   $\exp(31/66+\gamma)$ &$0.48811(1)$\\  
$b_{2}^{S}\times(4\pi)^3$   & & $1.6524$ & $2.1287$ & $4.3353$ & $0.483(9)$\\[1ex]
\hline 
  \end{tabular} 
\end{center}
  \caption{\label{t_lambda_b2} \footnotesize Ratio of $\Lambda$-parameters and 3-loop coefficient
           of the $\beta$-function for various schemes for $\nf=0$. These results follow from
           \protect\cite{Fischler:1977yf,Billoire:1980ih,MS:3loop1,MS:3loop3,Peter:1997me,Schroder:1998vy,Melles:2000dq,pert:2loop_nf0}.}
  \end{table} 
%%%%%%%%%%%%%%%%%%%%%%%%%%%%%%%%%%%%%%%%%%%%%%%%%%%%%%%%%%%%%%%%%%%%%%%%%%%%%%%%%%%%%%%%%%%%%

Note that the couplings in the
two schemes which are defined in terms of
the potential are restricted to perturbation theory.
Non-perturbatively it is not clear how to subtract the
self energy term in the potential and, in addition, performing
the Fourier transformation of numerical data known in
a finite range of $r$ is possible only in a model dependent way.
Also for this reason, $\alphaqqbar$ is the natural observable 
for the comparison between perturbation theory and
non-perturbative QCD.
\vspace{0.5cm}
%%%%%%%%%%%%%%%%%%%%%%%%%%%%%%%%%%%%%%%%%%%%%%%%%%%%%%%%%%%%%%%%%%%%%%%%%%%%%%%%%%%%%%%%
\begin{table}[ht]%\label{t_coeff}
\begin{center}
\begin{tabular}{ c|  c  c  c  c  c}
\hline\\[-1ex]
%$f_{2}^{\rm {SS'}}(s_{0})$ 
 $S'$ & $S\,=\,$  & $\MSbar$         & $\qqbar$  & $\V$     & $\Vbar$    \\[1ex]
\hline\\[-1ex]
$\qqbar$        & & $1.0653$ &              &              &                   \\
$\V$            & & $1.6095$ & $0.5441$ &               &                      \\
$\Vbar$         & & $4.1303$ & $3.0650$ &  $2.5208$   &                      \\
SF              & & $-0.271(10)$  & $-1.336(10)$  & $-1.880(10)$      & $-4.401(10)$             \\[1ex]
\hline
  \end{tabular} 
\end{center}
  \caption{\label{t_coeff} Coefficients $f_{2}^{ {S'S}}(s_{0})$ for $s_0=\Lambdasp/\Lambdas$ and $\nf=0$.}
  \end{table}
%%%%%%%%%%%%%%%%%%%%%%%%%%%%%%%%%%%%%%%%%%%%%%%%%%%%%%%%%%%%%%%%%%%%%%%%%%%%%%%%%%%%%%%%%%%

\ssect{3}
When two different couplings are known non-perturbatively, 
it is further of interest to study how well perturbation theory 
predicts their direct relation. This means matching the two 
couplings at finite $\mu$ instead of through the 
Lambda-parameter, which corresponds to matching for 
$\mu \to \infty$. 
The 
perturbative relation 
\be \label{e_match}
  \alphasp(s\mu) = \alphas(\mu) + f_1^{S'S}(s) \alphas(\mu)^2 + 
                      f_2^{S'S}(s) \alphas(\mu)^3 + \ldots  
\ee
contains a freedom of relative scale, $s$. 
Indeed, the choice of $s$ is in general very important for the 
quality of the perturbative prediction \cite{Peter:1997me,alpha:SU2impr}.
The only viable criterion for fixing $s$ appears to be to demand 
that the coefficients $f_i(s)$ are small (``fastest apparent convergence''). 
The choice $s=s_0=\Lambdasp/\Lambdas$ yields $f_1(s_0)=0$,
and in addition the values of $|f_2(s_0)|$ are close to the minimum 
of $|f_2(s)|$. 
The coefficients $f_2(s_0)$ connecting selected schemes 
are listed in \tab{t_coeff}. One observes that the SF-scheme is very close to
the $\MSbar$-scheme, the $\qqbar$ scheme is not very far, but the other 
schemes have quite large values of $f_2(s_0)$ in their relation to the 
$\MSbar$-scheme. In particular, the large coefficient between the $\MSbar$ 
scheme and
the $\Vbar$ scheme means that the direct expansion
of the coordinate space potential in terms of $\alphaMSbar$ (or $\alphaSF$)
 is badly behaved,
as it has been pointed out in \cite{Peter:1997me,Schroder:1999sg,Melles:2000dq}.
%%%%%%%%%%%%%%%%%%%%%%%%%%%%%%%%%%%%%%%%%%%%%%%%%%%%%%%%%%%%%%%%%%%%%%%%%%%%%%
\begin{figure}[ht]%\label{f_alpha_qq}
\begin{center}
\includegraphics[width=9cm]{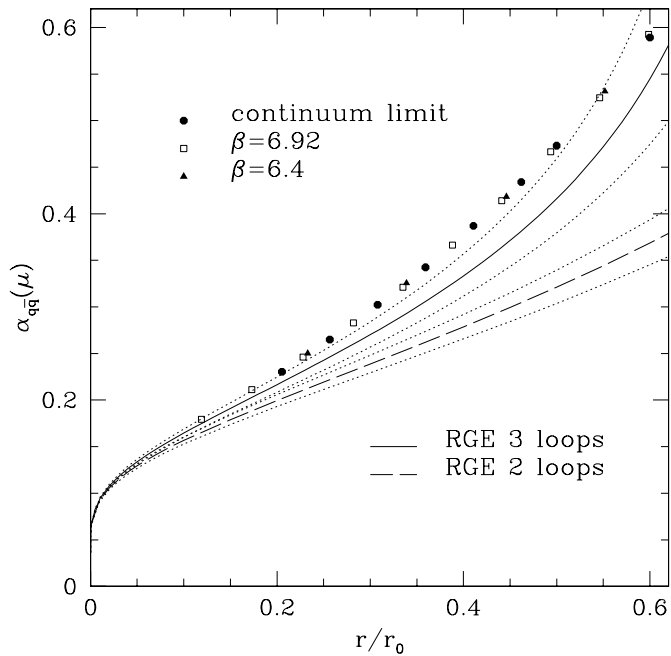}
\end{center}\vspace{-0.9cm}
\caption{\label{f_alpha_qq} \footnotesize
Running coupling in the $\qqbar$ scheme. Results for the continuum
limit as well as additional points at finite $\beta$, corresponding to finite
lattice spacing are shown. In the latter case the discretization errors were
estimated to be smaller than the size of the symbols.
The perturbative curves use $\Lambda_{\MSbar} \rnod$ from \protect\cite{mbar:pap1}
with the dotted lines corresponding to the 1-$\sigma$ uncertainties of
this combination.}
\end{figure}
%%%%%%%%%%%%%%%%%%%%%%%%%%%%%%%%%%%%%%%%%%%%%%%%%%%%%%%%%%%%%%%%%%%%%%%%%%%%%%

We emphasize the following point.  Although 
the three different schemes $\qqbar$, $V$, $\Vbar$ differ only by kinematics 
(differentiation, the Fourier transformation) it makes a big difference
for the applicability of perturbation theory which one is chosen to represent the
potential. The analysis of the perturbative series themselves  suggests
that potential and force should be reconstructed from $\alphaqqbar$.
\vspace{0.5cm}

\ssect{4}
 In \fig{f_alpha_qq} we compare the non-perturbative results of 
\cite{pot:silvia1} to perturbation theory. The 3-loop RG expression with 
$\Lambda_{\MSbar}$
at the upper end of the error bar of \eq{e_lamrnod}
is in very close agreement with the 
non-perturbative coupling. In fact the agreement extends up to values 
of $\alphaqqbar$ where perturbation theory is not
to be trusted a priory.
For $\alphaqqbar \lesssim 0.3$ our criterion (i) above
is satisfied but there is no evidence for 
non-perturbative terms in this region.

%%%%%%%%%%%%%%%%%%%%%%%%%%%%%%%%%%%%%%%%%%%%%%%%%%%%%%%%%%%%%%%%%%%%%%%%%%%%%%
\begin{figure}[ht]
\begin{center}
\includegraphics[width=10cm]{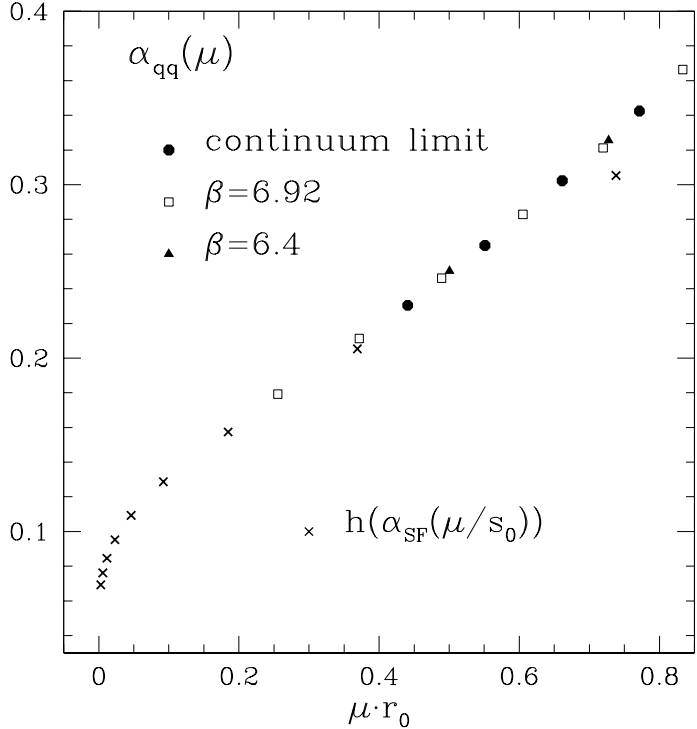}
\end{center}\vspace{-0.9cm}
\caption{\label{f_alphaqqsf}  \footnotesize
Test of \eq{e_alphaqqbar_pert}. The uncertainty in the combination $\mu \rnod$ 
has been 
translated into 
an uncertainty for $h(\alphaSF(\mu/s_0))$ and $\alphaqqbar(\mu)$.
The non-perturbative values for 
$\alphaSF(\mu)$ are constructed from the data of \protect\cite{mbar:pap1}. 
Errors are smaller than the sizes of the symbols.
}
\end{figure}
%%%%%%%%%%%%%%%%%%%%%%%%%%%%%%%%%%%%%%%%%%%%%%%%%%%%%%%%%%%%%%%%%%%%%%%%%%%%%%

The same conclusion is reached when we consider the relation
between $\alphaSF(\mu)$ and $\alphaqqbar(\mu)$ at finite $\mu$:
We define the 3-loop expression
\bes
 h(\alphaSF(\mu)) &=& \alphaSF(\mu)+ 1.336 \,[\alphaSF(\mu)]^3\, 
\ees
such that 
\bes
 \alphaqqbar(\mu) &=& h(\alphaSF(\mu/s_0)) + \rmO([\alphaSF(\mu/s_0)]^4)
\label{e_alphaqqbar_pert}
\ees
as explained above. Then the non-perturbative values of $\alphaqqbar(\mu)$
and of $h(\alphaSF(\mu/s_0))$ are compared 
in \fig{f_alphaqqsf}. If the higher order terms in 
\eq{e_alphaqqbar_pert} are negligible, the two different quantities should 
agree. 
At $\alphaqqbar(\mu)\approx 0.3$ a difference is visible
but this is only about $3\times \alpha^4$, not far from
the expected size of the next order term in the series.\footnote{
Note that the next order correction is formally enhanced by a 
logarithm of $\alpha$, which originates from a resummation
of IR divergent terms. It reads 
$(A\log(\alpha)+B)\alpha^4$ \cite{pot:logterms_1,pot:logterms_2,Schroder:1999sg}.
While $A=9/(4\pi)$ has recently been calculated
\cite{pot:logterms_3,pot:logterms_4}, $B$ is not known. 
The $A \alpha^4\log(\alpha)$ term by itself constitutes 
a small correction in the figure, which would slightly enlarge
the difference between $h(\alphaSF(\mu/s_0))$ and $\alphaqqbar(\mu)$.} 
At $\alphaqqbar(\mu) \approx 0.2$, the difference 
$\alphaqqbar(\mu) - h(\alphaSF(\mu/s_0))$ is not significant
at all. We conclude that also in \eq{e_alphaqqbar_pert}
a large non-perturbative term at short distances
is excluded.

%%%%%%%%%%%%%%%%%%%%%%%%%%%%%%%%%%%%%%%%%%%%%%%%%%%%%%%%%%%%%%%%%%%%%%%%%%%%%%
\begin{figure}[ht]
\begin{center}
\includegraphics[width=9cm]{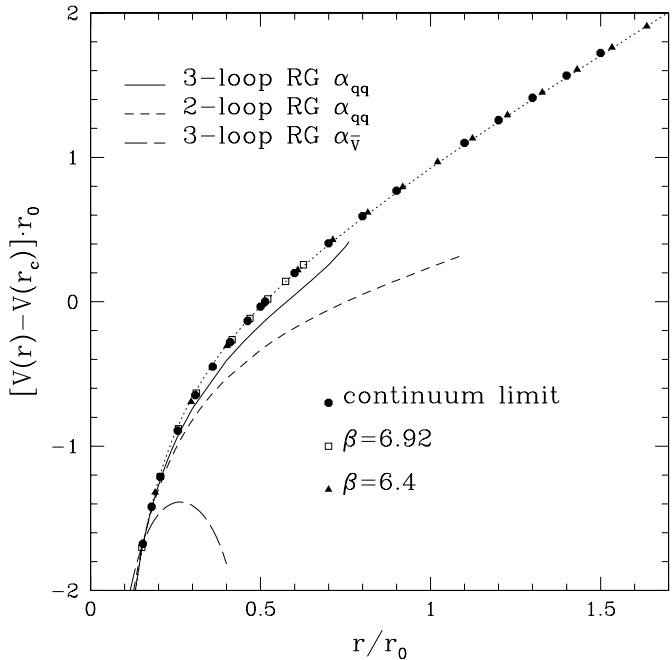}
\end{center}\vspace{-0.9cm}
\caption{\label{f_pot} 
The potential compared to different
perturbative expressions. Here, $r_{\rm c}=0.54\rnod$   \protect\cite{pot:silvia1}.
}
\end{figure}
%%%%%%%%%%%%%%%%%%%%%%%%%%%%%%%%%%%%%%%%%%%%%%%%%%%%%%%%%%%%%%%%%%%%%%%%%%%%%%

Finally we show in \fig{f_pot} the static potential itself \cite{pot:silvia1}
compared to different perturbative approximations. Full line and short dashes
are given by 
\bes\label{e_vint}
V(r) = V(0.15\rnod)+\int_{0.15\rnod}^r \rmd y F(y)\,,\;F(r)=\Cf r^2 \alphaqqbar(1/r)
\ees
with the 3-loop and 2-loop RG-solution for $\alphaqqbar$. Since we
know that the 3-loop RG-solution for $\alphaqqbar$ is accurate,
this also hold for $V(r)$ computed
through \eq{e_vint}. Again, the full line moves very close
to the data points ($r<0.5\rnod$), when $\Lambda \rnod$ at the upper end of the
error bar of \eq{e_lamrnod} is inserted. The long dashes represent \eq{e_alphavbar}
with the 3-loop RG-solution for $\alphavbar$. As it was to
be expected due to the missing
stability of this perturbative expression, it fails in describing the
potential. A similarly bad perturbative 
expression (not shown here)
is the direct expansion of the potential in terms of $\alphaMSbar$.

\vspace{0.5cm}

\ssect{5} In summary, care has to be taken which perturbative prediction (scheme)
is adopted to describe the potential. However, perturbation theory
does its best in the following sense. As usual in an asymptotic expansion,
one should first investigate the apparent ``convergence'' by comparing 
subsequent orders and checking that they decrease significantly.
If this is not the case, one is obviously outside the domain of
applicability of perturbation theory or has chosen a bad truncation (scheme).
According to this criterion the $\beta$-function in the 
$\qqbar$-scheme may be trusted up
to $\alphaqqbar\approx 0.3$.  Other
truncations of perturbation theory for the potential 
that we investigated are 
applicable for much smaller values of the coupling only. 
Therefore perturbation theory suggests that the $\qqbar$ scheme 
should be used
in order to obtain a reliable perturbative expression.\footnote{Of course,
other similarly well behaved truncations of  perturbation theory
might be found. The important point is that a scheme with a large
3-loop coefficient such as $\Vbar$ is of no use in the region
$\alpha > 0.15$.}

Our comparison
with non-perturbative results, obtained in the continuum
limit of lattice simulations
($\nf=0$), does confirm that such a  perturbative analysis is a good
guideline -- at least in the case at hand. Of course one should not 
expect miracles when one goes up to $\alpha\approx 0.3$. 
At such values of the coupling we only confirm that the 
3-loop perturbative 
prediction is good to within about 10\% and indeed in \fig{f_alphaqqsf}
one sees explicitly that the truncated  
perturbative series has {\em errors} of this order. Similar 
results have been found for the $\nf=2$ coupling in the SF-scheme
(see Fig. 5 in \cite{alpha:lett}).

Which lessons can we learn for QCD with quarks? Compared to $\nf=0$,
the relevant perturbative coefficients, $b_2$ and $f_2(s_0)$, which
are listed in
the appendix, are roughly a factor two smaller in magnitude
for $\nf=3$. This suggests that with quarks 
the perturbative prediction
for the potential computed through $\alphaqqbar$ is also applicable
up to $\alpha\approx 0.3$ 
and furthermore in full QCD the issue of the appropriate scheme
is somewhat less important. 
A direct lattice QCD check of these expectations 
is unfortunately not possible at present and here we 
had to boldly 
generalize from the $\nf=0$ case. In addition,
these remarks apply to the massless theory 
(we have not investigated mass effects). 
Current phenomenological research concentrates on the application
of a velocity dependent potential beyond the static limit 
for phenomenological applications to top-quark physics\cite{pot:Hoang:2000yr}. 
On the one hand,
in this application the potential is needed for quite short distances, 
where perturbation theory is intrinsically more precise \cite{topphys:kuehn},
on the other hand, with the velocity entering as a new scale,
this represents a more difficult multi-scale problem. Indeed the renormalization
group has already been applied to deal with this complication 
\cite{pot:Hoang:2001rr}. 
Nevertheless, the lessons learnt in our investigation
may be useful in this context as well; the type of renormalization
group improvement which we found to increase the reliability
of perturbation theory (see \fig{f_pot}) has not been applied in 
\cite{pot:Hoang:2000yr} so far.

\vspace{0.5cm}  
\ssect{Acknowledgements} We thank
F. Jegerlehner,
O. Tarasov and U. Wolff for useful discussions and N. Brambilla 
for correspondence. 
This work is supported by the European Community's Human potential 
programme under HPRN-CT-2000-00145 Hadrons/LatticeQCD.

\vspace{0.5cm}  
\ssect{Appendix} 
In order to ease the comparison of different schemes,
we here list some perturbative coefficients 
for general $\nf$ whose numerical values for $\nf=0$ were quoted
in the tables above. Some of them could be taken directly
from the literature 
\cite{Fischler:1977yf,Billoire:1980ih,MS:3loop1,MS:3loop3,Peter:1997me,Schroder:1998vy,Melles:2000dq,pert:2loop_fin},
others such as $b_2^{\qqbar}$ had to be
computed by straight forward algebra.

%%% Local Variables: 
%%% mode: latex
%%% TeX-master: "pap2"
%%% End: 

%%%%%%%%%%%%%%%%%%%%%%%%%%%%%%%%%%%%%%%%%%%%%%%%%%%%%%%%%%%%%%%%%%%%%%%%%%%%%%%%%%%%%%%%
\begin{table}[tb]
\begin{center}
\begin{tabular}{c| c| c| c }
\hline &&&\\[-1.9ex]
$S$ & $e_{1}$ & $e_{2}$ & $e_{3}$  \\[0.5ex]
\hline&&&\\[-1.9ex]
$\MSbar$ & $-{5033}/{18}$ & $325/54$  & $0$\\
$\rm{V}$      &  $\frac{3}{2}\pi^4-24\pi^2-\frac{2239}{6}-\frac{704}{3}\zeta(3)$ & $\frac{377}{54}+\frac{104}{9}\zeta(3)$ & $0$\\[0.5ex]
$\overline{\rm V}$ & $-\frac{2239}{6}-\frac{704}{3}\zeta(3)+\frac{3}{2}\pi^4-\frac{314}{3}\pi^2$ & $\frac{377}{54}+\frac{104}{9}\zeta(3)+\frac{44}{9}\pi^2$ & $-\frac{8}{81}\pi^2$\\[0.5ex]
$\rm{q\bar{q}}$  &  $\frac{3}{2}\pi^4-\frac{314}{3}\pi^2+\frac{3569}{6}-\frac{704}{3}\zeta(3)$ & $-\frac{2791}{54}+\frac{44}{9}\pi^2+\frac{104}{9}\zeta(3)$ & $\frac{32}{27}-\frac{8}{81}\pi^2$\\[0.5ex] 
$\rm{SF}$  & $-0.275(5)\times(4\pi)^{3}$ & $ 0.0361(4)\times(4\pi)^{3}$  
& $-0.00175(1)\times(4\pi)^{3}$ \\[0.5ex]
\hline
\end{tabular}
\end{center}
  \caption{\label{t_ei} \footnotesize Coefficients $e_i$ of \protect\eq{e_b2gen}.}
\end{table}
%%%%%%%%%%%%%%%%%%%%%%%%%%%%%%%%%%%%%%%%%%%%%%%%%%%%%%%%%%%%%%%%%%%%%%%%%%%%%%%%%%%%
The three-loop coefficient of the $\beta$-function can be expressed as
\be\label{e_b2gen}
b_{2}^{S}=b_{2}^{S}|_{\nf=0}+\frac{1}{(4\pi)^6}\left(e_{1}\nf+e_{2}\nf^2+e_{3}\nf^3\right)
\ee
with $e_i$ listed in \tab{t_ei}.

%%%%%%%%%%%%%%%%%%%%%%%%%%%%%%%%%%%%%%%%%%%%%%%%%%%%%%%%%%%%%%%%%%%%%%%%%%%%%%%%%%%%%%%%
\begin{table}[tb]
\begin{center}
\begin{tabular}{ c  c  c}
\hline\\[-1.9ex]
$S'$         & $a_{1}$ & $a_{2}$      \\[0.5ex]
\hline\\[-1.9ex]
$\rm V$              & $31/3$  &$-10/9$              \\[0.5ex]
$\overline{\rm V}$  & $31/3+22\gamma$ & $-10/9-4\gamma/3$  \\[0.5ex]
$\rm{q\bar q}$              & $-35/3+22\gamma$ & $2/9-4\gamma/3$ \\[0.5ex]
$\rm{SF}$   & $-1.25562 \times (4\pi)$   &  $-0.03986 \times (4\pi)$   \\[0.5ex]
\hline 
  \end{tabular} 
\end{center}
  \caption{\label{t_ai} \footnotesize Coefficients $a_i$ of \protect\eq{e_f1} for $S=\MSbar$}
  \end{table}
%%%%%%%%%%%%%%%%%%%%%%%%%%%%%%%%%%%%%%%%%%%%%%%%%%%%%%%%%%%%%%%%%%%%%%%%%%%%%%%%%%%%%%%%
The one-loop coefficient in \eq{e_match} assumes the general form
\be \label{e_f1}
f_{1}^{S'S}(s)=(a_{1}+a_{2}\nf)/4\pi- 8\pi b_{0}\log(s)),
\ee
and vanishes for $s=s_{0}=\exp\left((a_{1}+a_{2}\nf)/(32\pi^2 b_{0})\right)$.
$a_{1}$ and $a_{2}$ are listed in \tab{t_ai}, with $\MSbar$ 
as reference scheme. The other coefficients can be evaluated by
$f_{1}^{SS'}(s)=f_{1}^{SS''}(s)+f_{1}^{S''S'}(s)$ and
\be
f_{2}^{S'S}(s)=\frac{(4\pi)^2}{b_{0}}\left\{b_{2}^{S'}-b_{2}^{S}+b_{1}\frac{f_{1}^{S'S}(s)}{4\pi}-b_{0}\left[\frac{f_{1}^{S'S}(s)}{4\pi}^{2}\right]\right\}\,,
\ee
which reduces to
\be
f_{2}^{S'S}(s_{0})=\frac{(4\pi)^2}{b_{0}}\left[b_{2}^{S'}-b_{2}^{S}\right]
\ee
when we set $s=s_{0}$.

%%% Local Variables: 
%%% mode: latex
%%% TeX-master: "pap2"
%%% End: 

   \bibliography{pap2}        %or whatever your .bib file is
   \bibliographystyle{h-elsevier}   %if you use h-elsevier.bst
\end{document}